\documentclass{emulateapj}
\input{epsf}
\submitted{Accepted for publication in ApJ on March 21, 2012}
\shorttitle{The shapes of Milky Way satellites}

\shortauthors{E. L. {\L}okas et al.}

\begin{document}

\title{The shapes of Milky Way satellites: looking for signatures of tidal stirring}

\author{Ewa L. {\L}okas\altaffilmark{1},  Steven R. Majewski\altaffilmark{2}, Stelios Kazantzidis\altaffilmark{3},
Lucio Mayer\altaffilmark{4}, Jeffrey L. Carlin\altaffilmark{5},\\ David L. Nidever\altaffilmark{2}
and Leonidas A. Moustakas\altaffilmark{6}}

\altaffiltext{1}{Nicolaus Copernicus Astronomical Center, 00-716 Warsaw, Poland; lokas@camk.edu.pl}
\altaffiltext{2}{Department of Astronomy, University of Virginia, Charlottesville, VA 22904-4325, USA}
\altaffiltext{3}{Center for Cosmology and Astro-Particle Physics; and Department of Physics; and Department of Astronomy,
    The Ohio State University, Columbus, OH 43210, USA}
\altaffiltext{4}{Institute for Theoretical Physics, University of Z\"urich, CH-8057 Z\"urich, Switzerland}
\altaffiltext{5}{Department of Physics, Applied Physics, \& Astronomy, Rensselaer Polytechnic Institute, Troy,
	NY 12180-3590, USA}
\altaffiltext{6}{Jet Propulsion Laboratory, California Institute of Technology, Pasadena, CA 91109, USA}

\begin{abstract}
We study the shapes of Milky Way satellites in the context of the tidal stirring scenario for the formation
of dwarf spheroidal galaxies. The standard procedures used
to measure shapes involve smoothing and binning of data and thus may not be sufficient to detect
structural properties like bars, which are usually subtle in low surface brightness systems.
Taking advantage of the fact that in nearby dwarfs photometry of individual
stars is available we introduce discrete measures of shape based on the two-dimensional inertia tensor and
the Fourier bar mode. We apply these measures of shape first to a variety of simulated dwarf galaxies formed via tidal
stirring of disks embedded in dark matter halos and orbiting the Milky Way. In addition to strong mass loss and
randomization of stellar orbits, the disks undergo morphological transformation that typically
involves the formation of a triaxial bar after the first pericenter passage. These tidally induced bars
persist for a few Gyr before being shortened towards a more spherical shape if the tidal force is strong enough.
We test this prediction by measuring in a similar way the shape of nearby dwarf galaxies, satellites of the
Milky Way. We detect inner bars in Ursa Minor, Sagittarius, LMC and possibly Carina. In addition, six out of
eleven studied dwarfs show elongated stellar distributions in the outer parts that may signify transition to
tidal tails. We thus find the shapes of Milky Way satellites to be consistent with the predictions of the
tidal stirring model.
\end{abstract}


\section{Introduction}

The dwarf galaxies of the Local Group (for reviews see Mateo 1998, Tolstoy et al. 2009) display a variety of
shapes. While this is particularly true for dwarf irregular (dIrr) galaxies, it also applies to dwarf
spheroidals (dSph). Unlike globular clusters, they tend to exhibit significantly non-spherical images of
stellar content. The
origin of these shapes is a subject of ongoing debate but one of the interesting possibilities involves
the evolutionary connection between dIrrs and dSphs. This scenario proposes that dSphs evolved from the primordial
dIrrs via tidal stirring caused by the presence of a large host galaxy, such as the Milky Way or Andromeda
(Mayer et al. 2001).

The general picture painted by the tidal stirring model studied by the means of high resolution $N$-body
simulations (Mayer et al. 2001; Klimentowski et al. 2009; Kazantzidis et al. 2011)
involves strong mass loss and morphological transformation, as well as a rebuilding of the orbital
structure of the initial disk. In such simulations, a primordial disk galaxy embedded in an extended dark matter
halo is placed on different orbits around a Milky Way-sized galaxy and evolved for a significant fraction of the
Hubble time. A thorough study of different aspects of the tidal stirring scenario, for different initial dwarf
structures and different orbits, was recently performed by Kazantzidis et al. (2011) and {\L}okas et al. (2011).
An important and general feature of the morphological evolution of the dwarf established in these
studies is the formation and longevity of a triaxial, tidally induced bar. The bar, usually formed out of the
disk at first pericenter passage, tumbles with a period much shorter than the orbital period of the dwarf, which leads
to a number of observational biases in the characteristic parameters of the dwarfs ({\L}okas et al. 2011).
As the tidal evolution progresses, the bar shortens and the stellar component of the dwarf becomes more and more
spherical. If the tidal force is strong enough (on tight orbits or orbits with small pericenters), the final
shape of the stellar component can be almost exactly spherical.

In this paper we discuss selected observational characteristics of the tidal stirring model that may help to
distinguish it from other scenarios for the formation of dSph galaxies,
e.g., those in which they form in isolation by purely
baryonic processes such as cooling, star formation, feedback from supernovae
and UV background radiation (Ricotti \& Gnedin 2005; Tassis et al. 2008; Sawala et al. 2010).
If the overall picture suggested by the tidal stirring model is correct, some of the less evolved dwarfs
in the Local Group may still be in
the bar-like phase. Such elongated shapes are indeed observed at least in a few cases, such as Ursa Minor
(e.g., Irwin \& Hatzidimitriou 1995), Sagittarius (Majewski et al. 2003) and the more recently discovered
Hercules dwarf (Coleman et al. 2007).

\begin{table*}
\begin{center}
\caption{Properties of the simulated dwarfs. }
\begin{tabular}{llccccccccccl}
\hline
\hline
Simulation & Varied    & $r_{\rm apo}$ & $r_{\rm peri}$ & $T_{\rm orb}$ & $r_{1/2}$ &     & $1-B/A$   &           &
&  $A_2$    &     & Color in\\
           & parameter & [kpc]         & [kpc]          & [Gyr]         & [kpc]     & $x$ & $y$       & $z$       & $x$
& $y$       & $z$ & Figs.~\ref{cbbaapo},\ref{a2baapo}      \\
\hline
O1  & orbit                       &   125 &    25   & 2.09 & 0.45 & 0.20 & 0.31 & 0.15 & 0.14 & 0.23 & 0.09 & green   \\
O2  & orbit                       &\ \ 87 &    17   & 1.28 & 0.38 & 0.05 & 0.08 & 0.03 & 0.03 & 0.04 & 0.04 & red     \\
O3  & orbit                       &   250 &    50   & 5.40 & 0.68 & 0.69 & 0.74 & 0.15 & 0.56 & 0.64 & 0.13 & blue    \\
O4  & orbit                       &   125 &\ \ 12.5 & 1.81 & 0.32 & 0.12 & 0.32 & 0.23 & 0.06 & 0.15 & 0.11 & orange  \\
O5  & orbit                       &   125 &    50   & 2.50 & 0.64 & 0.58 & 0.62 & 0.10 & 0.51 & 0.55 & 0.07 & purple  \\
O6  & orbit                       &\ \ 80 &    50   & 1.70 & 0.60 & 0.36 & 0.47 & 0.18 & 0.30 & 0.43 & 0.16 & brown   \\
O7  & orbit                       &   250 &\ \ 12.5 & 4.55 & 0.51 & 0.34 & 0.50 & 0.25 & 0.22 & 0.41 & 0.21 & black   \\
S6  & $i(-45^{\circ})$            &   125 &    25   & 2.09 & 0.46 & 0.21 & 0.27 & 0.07 & 0.19 & 0.22 & 0.04 & green   \\
S7  & $i(+45^{\circ})$            &   125 &    25   & 2.09 & 0.52 & 0.12 & 0.28 & 0.17 & 0.07 & 0.22 & 0.16 & red     \\
S8  & $z_{\rm d}/R_{\rm d}(-0.1)$ &   125 &    25   & 2.09 & 0.41 & 0.26 & 0.35 & 0.13 & 0.18 & 0.29 & 0.11 & blue    \\
S9  & $z_{\rm d}/R_{\rm d}(+0.1)$ &   125 &    25   & 2.09 & 0.47 & 0.18 & 0.27 & 0.11 & 0.13 & 0.20 & 0.07 & orange  \\
S10 & $m_{\rm d}(-0.01)$          &   125 &    25   & 2.09 & 0.47 & 0.15 & 0.26 & 0.13 & 0.11 & 0.18 & 0.07 & purple  \\
S11 & $m_{\rm d}(+0.02)$          &   125 &    25   & 2.09 & 0.42 & 0.17 & 0.46 & 0.35 & 0.09 & 0.41 & 0.32 & magenta \\
S12 & $\lambda(-0.016)$           &   125 &    25   & 2.09 & 0.32 & 0.23 & 0.46 & 0.30 & 0.14 & 0.42 & 0.29 & cyan    \\
S13 & $\lambda(+0.026)$           &   125 &    25   & 2.09 & 0.60 & 0.07 & 0.13 & 0.06 & 0.04 & 0.09 & 0.04 & pink    \\
S14 & $c(-10)$                    &   125 &    25   & 2.10 & 0.45 & 0.05 & 0.09 & 0.04 & 0.06 & 0.07 & 0.01 & black   \\
S15 & $c(+20)$                    &   125 &    25   & 2.08 & 0.53 & 0.39 & 0.47 & 0.14 & 0.33 & 0.41 & 0.09 & gray    \\
S16 & $M_{\rm h}(\times 0.2)$     &   125 &    25   & 2.14 & 0.33 & 0.19 & 0.26 & 0.09 & 0.17 & 0.22 & 0.05 & brown   \\
S17 & $M_{\rm h}(\times 5)$       &   125 &    25   & 1.88 & 0.65 & 0.24 & 0.33 & 0.12 & 0.17 & 0.27 & 0.10 & yellow  \\
\hline
\label{properties}
\end{tabular}
\end{center}
\end{table*}
% !!!!!!!!!!!!!!!     rows for S12-13 were changed to S14-15 and vice versa to conform with the new notation

\begin{figure*}
\begin{center}
    \leavevmode
    \epsfxsize=17cm
    \epsfbox[15 0 525 130]{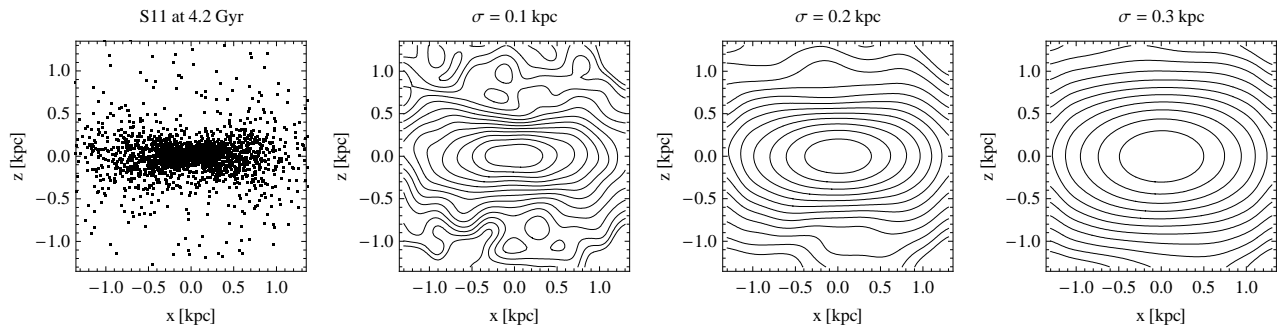}
\end{center}
\caption{The effect of smoothing on the detection of the bar-like shapes in dSph galaxies. The left panel shows
$2000$ stars randomly selected from all stars within $r<2$ kpc from the center of the dwarf at the
apocenter that occurred at 4.2 Gyr from the start of simulation S11
when the shape is most bar-like. The positions of the stars are
projected on the surface of the sky assuming that the dwarf is viewed along the intermediate ($y$) axis, i.e.,
perpendicular to the bar. The
remaining panels (from the left to the right) show the surface density of the stars measured in bins of 0.4
kpc on a side after application of Gaussian smoothing with standard deviation of $\sigma=0.1, 0.2$ and $0.3$
kpc respectively. The contours are plotted at constant numbers of stars $N$ with a step of $\log N = 0.2$ starting
with $\log N = 2$, 1.8 and 1.6 at the innermost contour for $\sigma=0.1$, 0.2 and 0.3 respectively.}
\label{smoothedbar}
\end{figure*}

The number of such detections may be low however for at least four reasons. First, the bars are oriented
randomly with respect to an observer located within the Milky Way, so some bar-like objects may appear as only
slightly flattened (which is indeed the case, given that the average ellipticity of dSph galaxies is about 0.3).
Second, the bars may escape detection due to the smoothing procedure usually applied when measuring the
surface density distribution of stars in these objects. Third, most dwarf galaxies are faint (thus containing
a low number of stars to sample) and observed against the
foreground of Milky Way stars, which can be only partially removed using typical cleaning procedures, such as
color-magnitude or color-color cuts. Fourth, the intrinsic bar-like shapes may be difficult to distinguish
from the tidal extensions in the surface distribution of the stars expected at close encounters between
the dwarfs and the larger host galaxy.

The measurements of the shapes of dwarf galaxies in the vicinity of the Milky Way are usually performed using
standard tools that involve binning and smoothing of the stellar distribution. In this work we try to improve
these procedures taking advantage of the fact that the satellites of the Milky Way are actually the best
extragalactic systems where individual stars can be detected and their spatial distribution studied. We introduce discrete
measures of the shape involving only the positions of the stars on the surface of the sky. We first apply these
procedures to simulated dwarfs to see what shapes are expected within the tidal stirring scenario. When
applying the procedures to real satellites of the Milky Way we use the cleanest possible stellar samples, selected using
the information provided by color-magnitude as well as color-color diagrams for the stellar populations.

The paper is organized as follows. In section 2 we briefly describe the simulations used in this study. In section 3 we
present our motivation for using the discrete measures of shape and introduce them. Section 4 presents the
application of these measures to the simulated dwarf galaxies formed via tidal stirring for a variety of
initial configurations. In section 5 we measure the shapes of the real dwarf galaxies
found in the vicinity of the Milky Way in an analogous way and compare the results to predictions from simulations.
The discussion follows in section 6.

\begin{figure}
\begin{center}
    \leavevmode
    \epsfxsize=8.3cm
    \epsfbox[60 0 310 220]{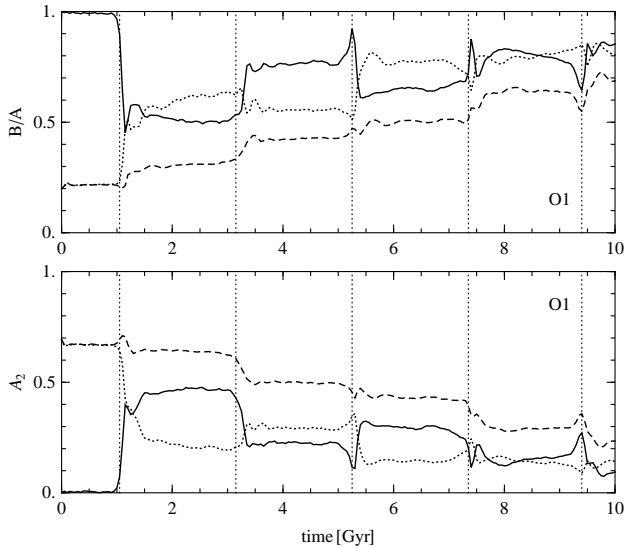}
\end{center}
\caption{The shapes of the stellar component in simulation O1 as a function of time. The upper panel
shows the axis ratios $b/a$ (solid line), $c/a$ (dashed line) and $c/b$ (dotted line), which correspond to the
line-of-sight axis ratio $B/A$ when the dwarf is observed along the $z$, $y$ and $x$ axis respectively.
The lower panel shows the measurements of the bar parameter $A_2$ when the dwarf is seen along the $z$ axis
(solid line) $y$ axis (dashed line) and $x$ axis (dotted line) as a function of time for simulation O1.
Vertical dotted lines indicate pericenter passages. The principal axes and the shape values were
determined for stars within $2 r_{1/2}$, where $r_{1/2}$ is the 3D half-light radius determined at the end
of the simulation. The final values of the shape parameters are listed in Table~\ref{properties}.}
\label{shapesrun1}
\end{figure}

\section{The simulations}

In this work we use the set of collisionless $N$-body
simulations presented in detail by Kazantzidis et al. (2011) and {\L}okas et al. (2011). The
purpose of these studies was to elucidate the formation of dSph galaxies via tidal interactions between late-type,
rotationally supported dwarfs and Milky Way-sized hosts.
The simulations, together with all the parameters relevant for the present study, are listed in Table~\ref{properties}
with the notation following that of {\L}okas et al. (2011). The simulations were performed using the
multistepping, parallel, tree $N$-body code PKDGRAV (Stadel 2001).

Numerical realizations of dwarf galaxy models were constructed using the method of Widrow \& Dubinski (2005).
The dwarfs were composed of exponential stellar disks embedded in cuspy,
cosmologically motivated Navarro et al. (1996, hereafter NFW) dark
matter halos. The default dwarf galaxy model had a virial mass
of $M_{\rm h} =10^9$ M$_{\odot}$ and a concentration parameter $c=20$.
The disk mass fraction, $m_{\rm d}$, and the halo spin
parameter, $\lambda$, were equal to $0.02$ and $0.04$,
respectively. The resulting disk radial scale length was $R_{\rm d} =
0.41$ kpc (Mo et al. 1998) and the disk thickness was specified by the
thickness parameter $z_{\rm d}/R_{\rm d} = 0.2$, where $z_{\rm d}$ denotes the
(sech$^2$) vertical scale height of the disk. The orientation of the internal angular momentum
of the dwarf with respect to the orbital angular momentum was mildly prograde
and equal to $i=45^{\circ}$.
Each dwarf galaxy model contained $N_{\rm h} = 10^6$ dark matter particles and $N_{\rm d} = 1.2 \times 10^6$ disk
particles. The gravitational softening was set to $\epsilon_{\rm h}=60$~pc
and $\epsilon_{\rm d}=15$~pc for the particles in the two components,
respectively.

\begin{figure*}
\begin{center}
    \leavevmode
    \epsfxsize=17.2cm
    \epsfbox[20 0 510 465]{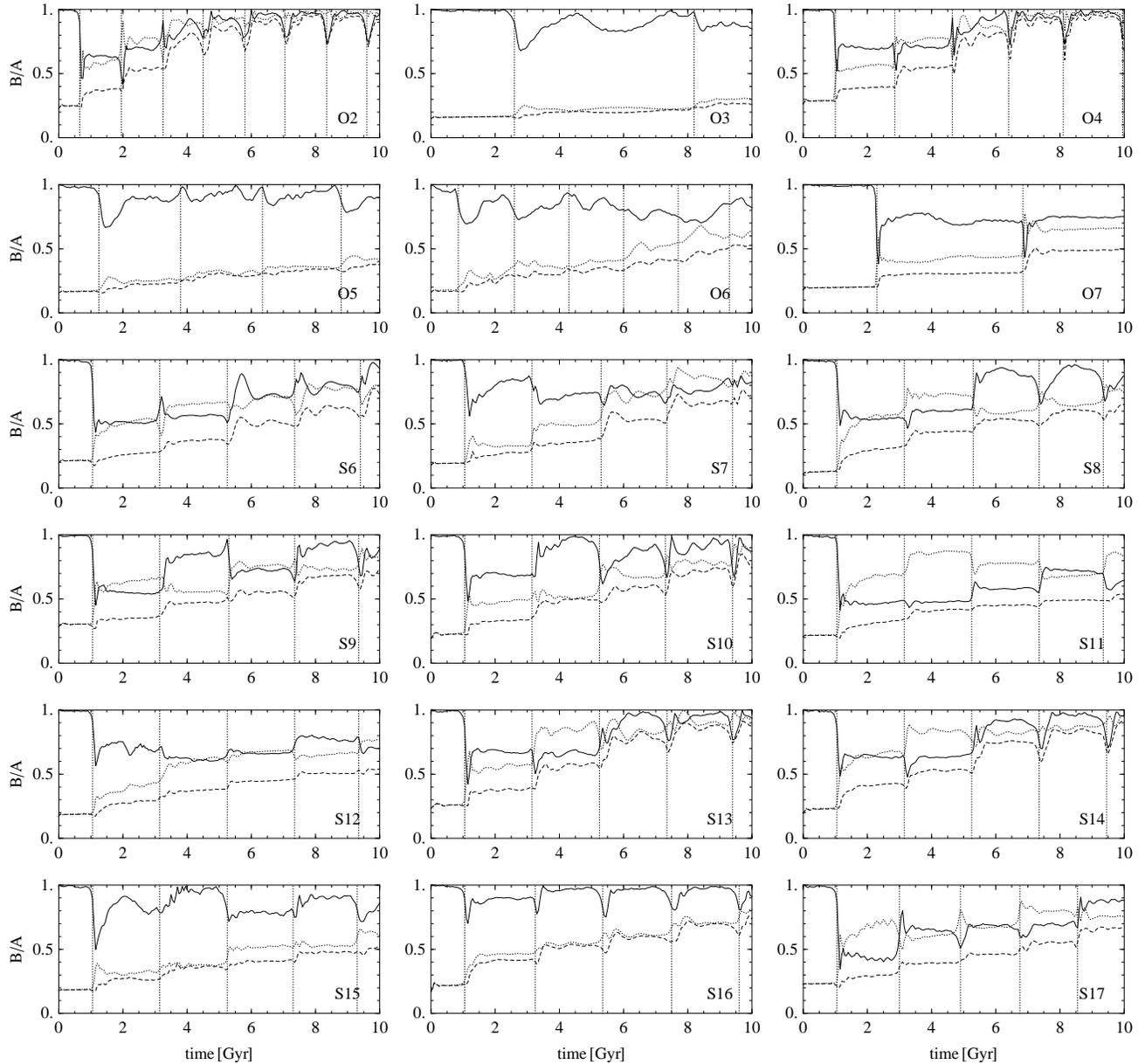}
\end{center}
\caption{The axis ratios $b/a$ (solid line), $c/a$ (dashed line) and $c/b$ (dotted line) for simulations O2-S17.
The values correspond to the
line-of-sight axis ratio $B/A$ when the dwarf is observed along the $z$, $y$ and $x$ axis respectively.
The values were determined for stars within $2 r_{1/2}$ with $r_{1/2}$ adopted to be the final value for
each simulation. Vertical dotted lines indicate pericenter passages.
The final values of the shape parameters are listed in Table~\ref{properties}.}
\label{axisratios2-19}
\end{figure*}

\begin{figure*}
\begin{center}
    \leavevmode
    \epsfxsize=17.2cm
    \epsfbox[20 0 510 465]{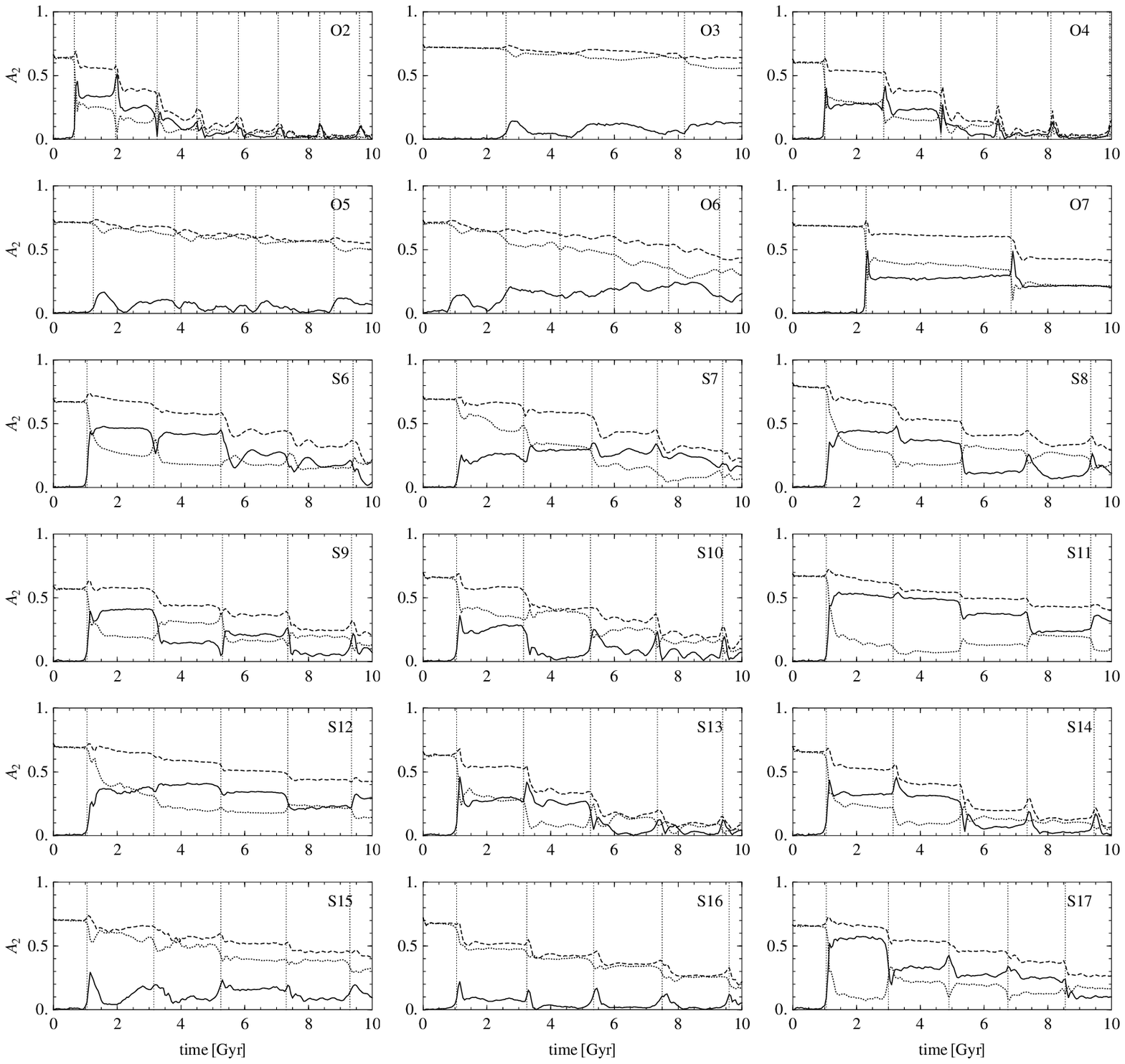}
\end{center}
\caption{The measurements of the bar parameter $A_2$ when the dwarf is seen along the $z$ axis
(solid line) $y$ axis (dashed line) and $x$ axis (dotted line) as a function of time for simulations O2-S17.
The values were determined for stars within $2 r_{1/2}$ with $r_{1/2}$ adopted to be the final value for
each simulation. Vertical dotted lines indicate pericenter passages.
The final values of the shape parameters are listed in Table~\ref{properties}.}
\label{a2xyz2-19}
\end{figure*}

The Milky Way model was
constructed as a live realization of the MWb model in Widrow \&
Dubinski (2005), which satisfies a broad range of observational
constraints for the present day Galaxy and consists of an exponential stellar
disk of mass $M_{\rm D}=3.53 \times 10^{10}$ M$_{\odot}$, a Hernquist (1990) bulge
of mass $M_{\rm B}=1.18 \times 10^{10}$ M$_{\odot}$, and an NFW dark matter halo
of mass $M_{\rm H}=7.35 \times 10^{11}$ M$_{\odot}$.
The $N$-body realization of the Milky Way used $N_{\rm D}=10^{6}$ particles
in the disk, $N_{\rm B}=5\times10^{5}$ in the bulge, and $N_{\rm H}=2\times10^{6}$
in the dark matter halo and employed a gravitational
softening of $\epsilon_{\rm D}=50$~pc, $\epsilon_{\rm B}=50$~pc, and
$\epsilon_{\rm H}=2$~kpc, respectively.

The default dwarf galaxy model was placed on seven different orbits (O1-O7 in Table~\ref{properties})
around the primary galaxy with orbital apocenters, $r_{\rm apo}$, and pericenters, $r_{\rm peri}$,
listed in the third and fourth column of Table~\ref{properties}.
In simulations S6-S17 (see Table~\ref{properties}), we varied the orientation of the dwarf galaxy disk and the
structural parameters of the dwarfs while keeping them on the same
orbit (that used for run O1, with $r_{\rm apo} = 125$ kpc and $r_{\rm peri} = 25$ kpc). The
second column of Table~\ref{properties} lists the parameters that
were varied in each case (in parentheses we give the value by which a
given parameter was changed). In all simulations, the dwarfs were
initially placed at the apocenters of their orbits and the evolution was followed for $10$~Gyr.
The orbital times corresponding to these orbits are listed in the fifth column of the Table.

\section{Measuring shapes}

The usual approaches to measuring shapes of the Local Group dwarf galaxies involve procedures such as binning
and smoothing of the stellar positions and fitting ellipses to the 2D distribution of stars. The results of such
procedures are typically quantified in terms of the ellipticity (or axis ratio) and major axis orientation
as a function of radius. For most dwarfs, a single ellipticity parameter, $e=1-b/a$ (with $a$ and $b$ the major and
minor axis length respectively) is measured (e.g., Mateo 1998), primarily in the outer parts of the dwarf.
When the number of stars available is low, their positions are often smoothed by applying a Gaussian smoothing.
If the smoothing scale is not carefully chosen, such procedures may cause the dwarfs to appear
more circular than they really are, removing substructures such as bars, or at least making them less pronounced.

We illustrate this effect in Figure~\ref{smoothedbar} by performing a similar procedure on one of our simulated dwarfs.
For this purpose we selected stars within radius $r<2$ kpc from the center of the dwarf (so that there is no
contamination from the tidal tails) from the output of simulation S11 when the dwarf is at
apocenter (after 4.2 Gyr from the start of the simulation) and its shape is most bar-like.
From this sample we randomly chose 2000 stars whose positions (plotted as dots in the left panel of
Figure~\ref{smoothedbar}) were projected on a plane perpendicular to the line of sight. The line of sight was chosen
to be perpendicular to the longest and shortest axis of the bar where the detection of the bar should be easiest.

We then replaced each star by a 2D Gaussian centered on
the star and normalized to unity with a standard deviation $\sigma$. The observed region was divided into bins
of $0.4 \times 0.4$ kpc and the number of stars in each bin was calculated by summing up the contributions from all
Gaussians. The resulting surface density distributions of the stars obtained with $\sigma=0.1, 0.2$ and $0.3$ kpc
are shown respectively in the second, third and fourth panel of Figure~\ref{smoothedbar}. We can see that while with
$\sigma=0.1$ kpc the bar is quite well visible, with $\sigma=0.3$ kpc the shape would hardly be classified as
bar-like and more probably described as highly elliptical. Note that the values of the smoothing parameter $\sigma$
adopted here are quite realistic, i.e., similar to what is usually applied to real data. For example, in
studies of the Leo I dSph galaxy Sohn et al. (2007) and Mateo et al. (2008) used $\sigma=1.1$ and $2$
arcmin respectively, which corresponds to 0.2 and 0.4 kpc, whereas the half-light radius of Leo I is only 0.25 kpc.
We learn from this exercise that when the smoothing scale becomes comparable to the half-light radius of the
dwarf ($r_{1/2}=0.48$ kpc for our simulated dwarf at the selected time), interesting morphological features
such as bars may no longer be detectable even if the line of sight is perpendicular to the bar.

\begin{figure}
\begin{center}
    \leavevmode
    \epsfxsize=7cm
    \epsfbox[10 10 215 420]{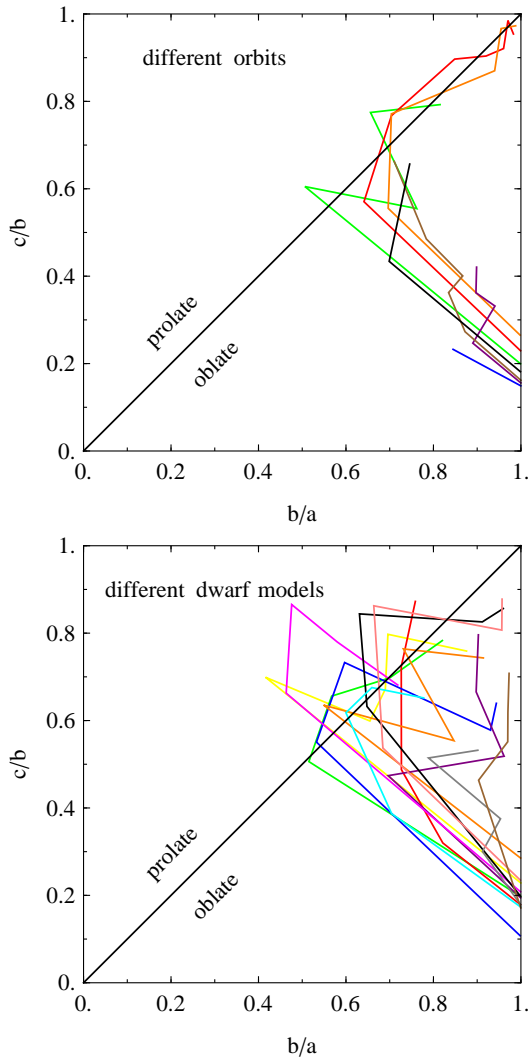}
\end{center}
\caption{The evolution of the axis ratio $c/b$ versus $b/a$ for dwarfs on different orbits O1-O7 (upper panel)
and dwarfs with different initial structure S6-S17 (lower panel). Color lines join measurements at subsequent apocenters
starting from the initial values at the lower right region
of each panel. The diagonal black line marks the transition between the oblate ($b/a > c/b$) and prolate
($b/a < c/b$) regimes.}
\label{cbbaapo}
\end{figure}

To avoid such problems, in this study we will quantify shapes using only {\em discrete\/} measures,
i.e., ones that use the 2D positions of the stars, without applying any smoothing or binning procedures.
We also aim to measure the {\em intrinsic\/} shapes, i.e., in the absence of any contamination from the
Milky Way or tidally stripped stars. Although avoiding such contamination is not possible in real observations, when
talking about our comparison shapes from the simulations, we will always measure such intrinsic shapes using
stars found within $2 r_{1/2}$ from the center of the simulated dwarf, where $r_{1/2}$ is the 3D half-light
radius containing half the total number of stars in 3D at the end of the simulation, as discussed in
{\L}okas et al. (2011). The values of these radii for our models are listed in the sixth column of Table~\ref{properties}.
These fixed radii are used for all outputs and all lines of sight whenever the shapes of the simulated dwarfs are
discussed. Note that such $r_{1/2}$ values are different from their 2D counterparts that in general depend
on the line of sight.

We will use two measures of the shape, both calculated from the 2D distribution of stars within a fixed radius.
In each case we first determine the principal axes of the stellar distribution from the 2D inertia tensor
and rotate the stars so that their $x(y)$ positions lie along the major (minor) axis. The first measure of
the shape we apply is the axis ratio $B/A$, or alternatively the ellipticity $e = 1 - B/A$, where
$B/A = [\Sigma_i y_i^2/\Sigma_i x_i^2]^{1/2}$. The second measure of the shape we use is the
so-called bar amplitude of the Fourier decomposition of stellar phases,
$A_2 = |\Sigma_{j} \exp(2 i \phi_j)|/N$ where $\phi_j$ are the phases of the positions
of stellar particles in the $xy$ plane. As we will demonstrate below, these measures, although slightly different in
their numerical values, are highly correlated and can be used alternatively.

\begin{figure*}
\begin{center}
    \leavevmode
    \epsfxsize=16cm
    \epsfbox[15 0 530 390]{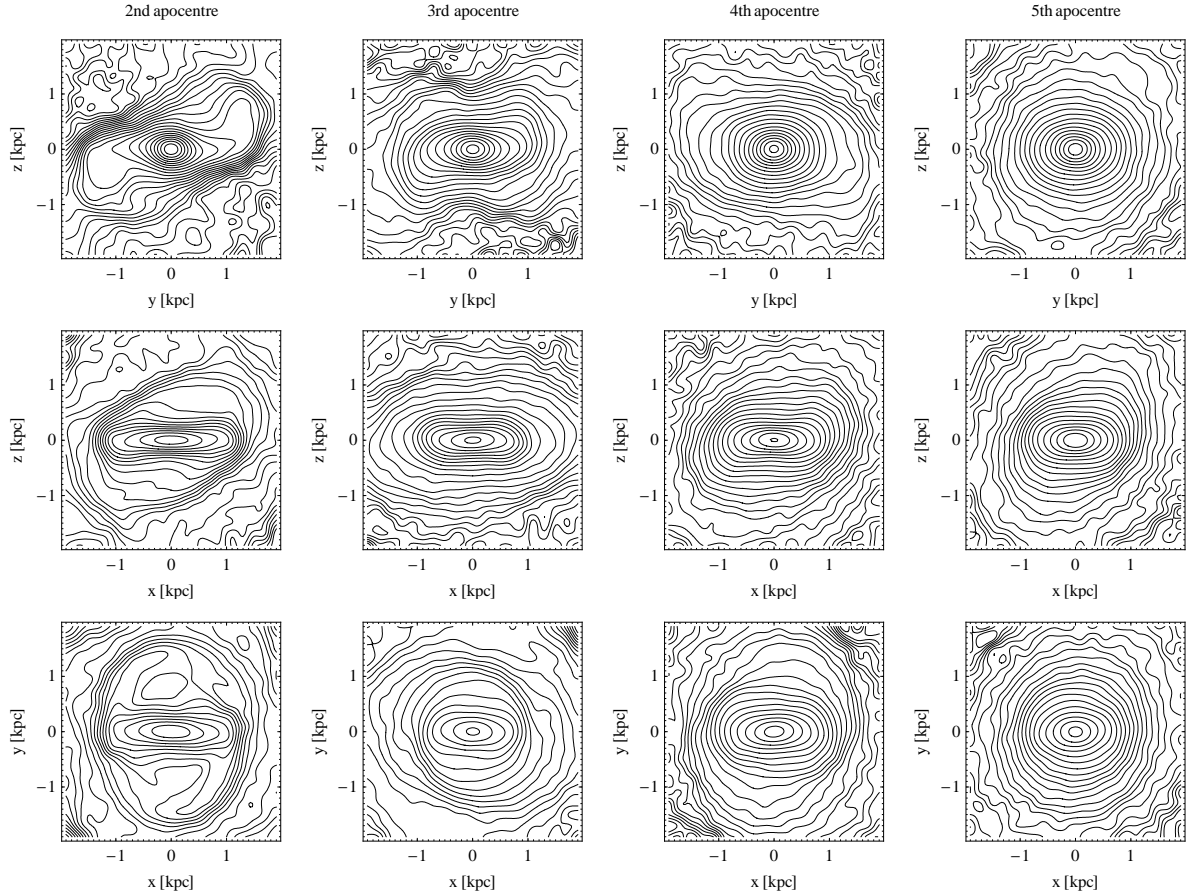}
\end{center}
\caption{The surface density distribution of the stars seen in simulation O1 at the second, third, fourth and fifth apocenter
(from the left to the right column) along the principal axes of the stellar distribution $x$, $y$ and $z$
(from the upper to the lower row). The bar formed after the first pericenter, positioned along the $x$ axis,
is gradually diminished at later stages.}
\label{viewsrun1apo}
\end{figure*}

\section{The shapes of the simulated dwarfs}

To measure the shapes of the stellar components of the simulated dwarfs, for each output of each run listed in the
first column of Table~\ref{properties} we selected stars contained within $2 r_{1/2}$ from the dwarf's center. The
3D half-light radius $r_{1/2}$ was kept at a fixed value for all outputs of a given run, and equal to its final value
at the end of the simulation, i.e., after 10 Gyr of evolution. Although $r_{1/2}$ depends slightly on time
(see {\L}okas et al. 2011), by keeping the radius fixed we make sure that the shapes are measured at the same scale for
all outputs and therefore that their evolution in time is easier to interpret (note that in Kazantzidis et al. 2011 the
shapes for the same runs were determined within a variable radius $r_{\rm max}$ corresponding to the maximum
circular velocity $V_{\rm max}$).

For each sample of stars selected in this way, we determine the orientation of the principal axes from the 3D
inertia tensor and rotate the distribution to align stellar positions with those axes, so that $x$ is along the
longest axis $a$, $y$ along the intermediate axis $b$ and $z$ along the shortest axis $c$. The 3D shapes of the dwarfs
can then be specified by the axis ratios
$b/a = [\Sigma_i y_i^2/\Sigma_i x_i^2]^{1/2}$,
$c/a = [\Sigma_i z_i^2/\Sigma_i x_i^2]^{1/2}$ and
$c/b = [\Sigma_i z_i^2/\Sigma_i y_i^2]^{1/2}$
where $(x_i, y_i, z_i)$ is the position of the $i$th star.
The 2D shapes are then obtained by selecting a line of sight. If the lines of sight are along the $z$, $y$ and $x$ axis,
then the measured 2D axis ratio $B/A$ will be equal to $b/a$, $c/a$ and $c/b$ respectively. An example of the shapes
measured in this way is shown as a function of time in the upper panel of Figure~\ref{shapesrun1} for simulation O1.
Using the same lines
of sight we measure the bar mode parameter $A_2$ which is shown in the lower panel of this figure. Analogous results
for the remaining runs O2-O7 and S6-S17 are plotted in Figures~\ref{axisratios2-19} and \ref{a2xyz2-19}.

The results presented in Figures~\ref{shapesrun1}-\ref{a2xyz2-19} confirm those obtained in a slightly different
way by Kazantzidis et al. (2011): in all cases after the first pericenter the stellar component transforms from
a disk to a triaxial ellipsoid with $a \neq b \neq c$. In most runs, at some point in the evolution, the shape becomes
prolate with $c/b > b/a$ which can be interpreted as the formation of a bar-like shape. This is confirmed by high
values of $A_2 > 0.2$ detected for the line of sight along the $z$ axis, i.e. signifying the formation of a bar in
the plane of the initial disk. As the evolution progresses, the shape becomes more spherical with all axis ratios
approaching unity, with the most significant changes occurring at the pericenter passages (vertical dotted lines
in Figures~\ref{shapesrun1}-\ref{a2xyz2-19}) when the dwarfs are most strongly tidally shocked.

The evolution of the global shape of the stellar component can be illustrated further by plotting the evolutionary
tracks of the dwarfs in the $b/a-c/b$ plane. We show them in Figure~\ref{cbbaapo} with lines of different color for
each simulation (see the last column of Table~\ref{properties}) for runs O1-O7 in the upper panel and runs
S6-S17 in the lower one. The tracks start at the lower right corner of each panel and progress towards the upper left
in subsequent apocenters. Most of the tracks cross the diagonal black line at some point, signifying the transition
from the oblate to the prolate shape, which can be interpreted as the formation of a bar. These correspond to the
dotted line being above the solid line in the upper panel of Figure~\ref{shapesrun1} and in Figure~\ref{axisratios2-19}.
Interestingly, all transitions into the prolate half of the diagram are accompanied by significantly high values
of $A_2$ measured along the shortest ($z$) axis (solid line) in the lower panel of Figure~\ref{shapesrun1}
and in Figure~\ref{a2xyz2-19}. This is due to the fact that if the bar mode occurs in the initial disk plane usually
it means that the global shape of the dwarf becomes bar-like, i.e.,
most stars participate in the bar instability and form a bar
while very few stars remain elsewhere. On the other hand, in cases when the shape remains oblate
during the whole evolution (runs O3, O5, O6, O7, S10, S15 and S16) the $z$-axis bar mode remains low with $A_2 < 0.2$.
An interesting borderline case is S10 when a bar mode is weakly excited ($A_2 \approx 0.2$ between the second and third
pericenter) but the overall shape of the dwarf remains oblate ($c/b < b/a$ at all apocenters).

\begin{figure}
\begin{center}
    \leavevmode
    \epsfxsize=7cm
    \epsfbox[0 0 240 245]{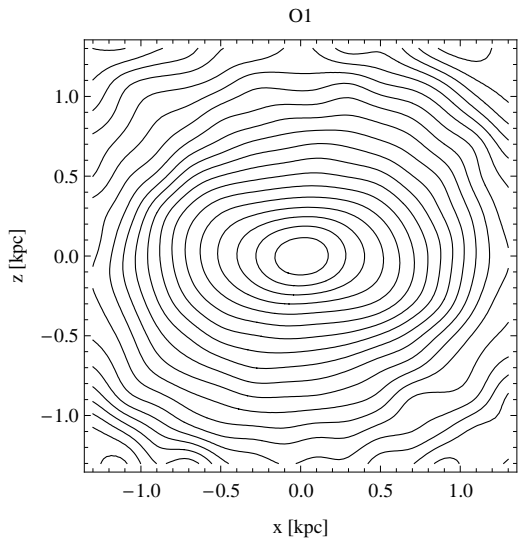}
\end{center}
\caption{The surface density distribution of the stars seen at the final output of simulation O1 along the $y$
(intermediate) axis of the stellar distribution, i.e., at its most non-spherical appearance.}
\label{shapealongyrun1}
\end{figure}

In addition to the $z$-axis bar mode shown in the lower panel of Figure~\ref{shapesrun1} and in Figure~\ref{a2xyz2-19}
with solid lines, we also add (with dashed and dotted lines respectively) the analogous measurements of this
parameter done when the dwarf is seen along the $y$ and $x$ axis of the stellar distribution. Obviously, a distant
observer will see the dwarf along a random line of sight and will not be able to determine a full 3D shape
of the stellar component. We thus show what such measurements would give in terms of $A_2$ if the line of sight
was close to the $y$ or $x$ axis. In these cases, the dwarf may often appear elongated not because it is a bar but because it
is still a disk, as demonstrated by high values of $A_2$ at the early stages of evolution. Our observer may not be
able to distinguish a disk from a bar even if kinematical measurements were available because, according to our simulations,
a tidally induced bar is expected to tumble (see Figure 4 in {\L}okas et al. 2011)
and thus may display a similar rotation curve as a disk. Only additional precise
photometric measurements aiming to constrain the depth of the dwarf galaxy could differentiate between
a disk and bar in this case.

\begin{figure*}
\begin{center}
    \leavevmode
    \epsfxsize=11cm
    \epsfbox[15 0 410 800]{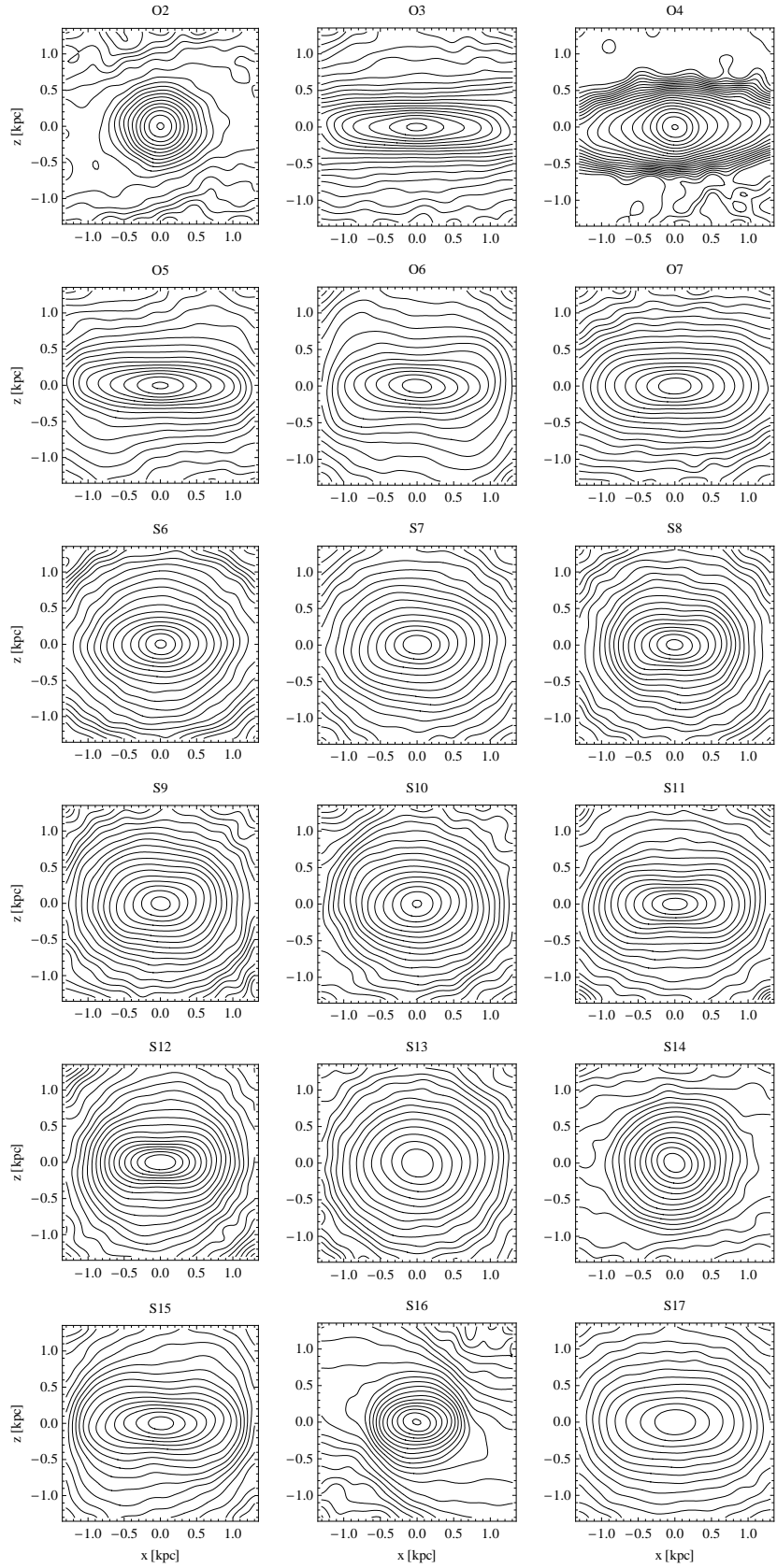}
\end{center}
\caption{The surface density distribution of the stars seen at the final output of simulations O2-S17 along the $y$
(intermediate) axis of the stellar distribution, i.e., at their most non-spherical appearance. For all simulations
the distributions are shown for stars within $r < 2$ kpc although the orientation of the principal axes of the
stellar component were determined using stars within $r < 2 r_{1/2}$ with $r_{1/2}$ different for each simulation.
The stellar distribution in O4 stands out with its pronounced, elongated outer contours due to the
very recent pericenter passage.}
\label{shapesalongy}
\end{figure*}

\begin{figure*}
\begin{center}
    \leavevmode
    \epsfxsize=17cm
    \epsfbox[60 0 540 145]{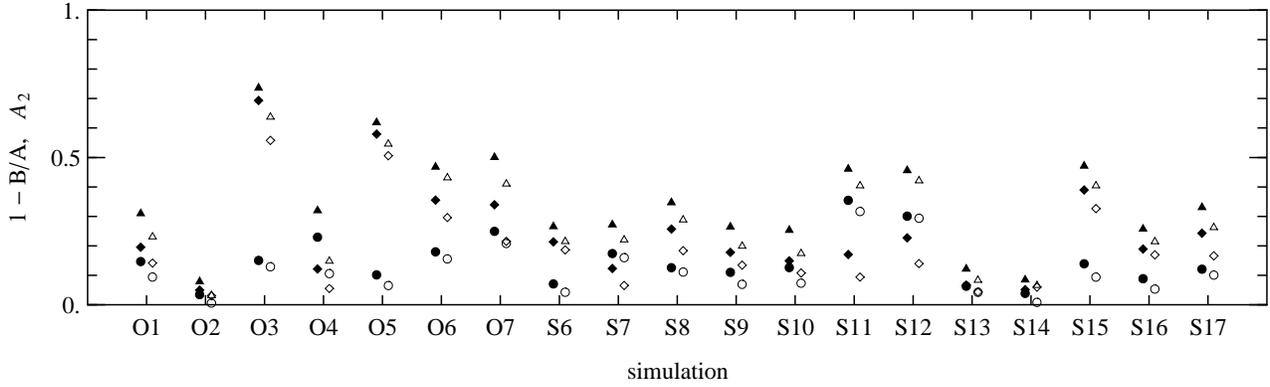}
\end{center}
\caption{The line-of-sight shape parameters $1-B/A$ (filled symbols) and $A_2$ (open symbols) determined from
the 2D surface density distribution of the stars in the simulated dwarfs.
The triangles, squares and circles refer to measurements performed
for observations along the intermediate ($y$), longest ($x$) and shortest ($z$) axis of the stellar distribution.
The plotted values are listed in Table~\ref{properties}.}
\label{losshapes}
\end{figure*}

\begin{table*}
\caption{Properties of the Milky Way satellites included in this analysis.}
\label{dwarfs}
\begin{center}
\begin{tabular}{lccrccrcl}
\hline
\hline
Dwarf   & Center  	  & Center          &  $d_{\odot}$\ \ & $r_{1/2}$ & $r_{1/2}$ & $N_{\rm stars}\ \ $  & Selection & Source   \\
galaxy  & $\alpha$(J2000) & $\delta$(J2000) & [kpc]        & [arcmin]  & [kpc]     & $(< 2 r_{1/2})$  & method    & of data  \\
\hline
Fornax     & $02^{\rm h}39^{\rm m}48^{\rm s}$ & $-34^{\circ}31'01''$ & 138 & 16.3  & 0.656 &  6283   & CM         & 1,2,1    \\
Sculptor   & $01^{\rm h}00^{\rm m}06^{\rm s}$ & $-33^{\circ}42'40''$ &  86 &  8.85 & 0.221 &  1751   & CM+CC      & 3,4,3    \\
Carina     & $06^{\rm h}41^{\rm m}37^{\rm s}$ & $-50^{\circ}58'00''$ & 105 &  7.86 & 0.240 &   269   & CM+CC      & 5,6,7    \\
Ursa Minor & $15^{\rm h}09^{\rm m}04^{\rm s}$ & $+67^{\circ}13'51''$ &  76 & 14.4  & 0.318 &   465   & CM+CC      & 8,9,10   \\
Sextans    & $10^{\rm h}13^{\rm m}03^{\rm s}$ & $-01^{\circ}36'54''$ &  95 & 27.9  & 0.770 &  2503   & CM         & 5,11,12  \\
Draco      & $17^{\rm h}20^{\rm m}19^{\rm s}$ & $+57^{\circ}54'48''$ &  80 &  9.66 & 0.225 &  1466   & CM         & 5,13,12  \\
Leo I      & $10^{\rm h}08^{\rm m}29^{\rm s}$ & $+12^{\circ}18'20''$ & 257 &  3.34 & 0.250 &  1100   & CM+CC      & 14,14,14 \\
Leo II     & $11^{\rm h}13^{\rm m}29^{\rm s}$ & $+22^{\circ}09'06''$ & 233 &  2.28 & 0.155 &   272   & CM+CC      & 15,16,17 \\
Sagittarius& $18^{\rm h}54^{\rm m}59^{\rm s}$ & $-30^{\circ}27'38''$ &  25 &  221  & 1.604 &  3344   & CM+CC      & 18,19,18 \\
LMC        & $05^{\rm h}23^{\rm m}34^{\rm s}$ & $-69^{\circ}45'24''$ &  50 &  136  & 1.979 & 132549  & CM         & 5,20,21  \\
SMC        & $00^{\rm h}51^{\rm m}00^{\rm s}$ & $-73^{\circ}07'12''$ &  62 & 70.9  & 1.278 &  13626  & CM         & 22,23,21 \\
\hline
\end{tabular}
\end{center}
\tablerefs{
(1) Coleman et al. (2005);
(2) Saviane et al. (2000);
(3) Westfall et al. (2006);
(4) Pietrzy\'{n}ski et al. (2008);
(5) Mateo (1998);
(6) Pietrzy\'{n}ski et al. (2009);
(7) Majewski et al. (2000b);
(8) Kleyna et al. (1998);
(9) Bellazzini et al. (2002);
(10) Palma et al. (2003);
(11) Lee et al. (2003);
(12) SDSS DR8, Aihara et al. (2011);
(13) Aparicio et al. (2001);
(14) Sohn et al. (2007);
(15) Coleman et al. (2007);
(16) Bellazzini et al. (2005);
(17) Siegel \& Majewski (2000);
(18) Majewski et al. (2003);
(19) Kunder \& Chaboyer (2009);
(20) Schaefer (2008);
(21) 2MASS survey, Skrutskie et al. (2006);
(22) Gonidakis et al. (2009);
(23) Szewczyk et al. (2009).}
\end{table*}

\begin{table*}
\caption{The shape parameters of Milky Way satellites included in this analysis.}
\label{shapeparameters}
\begin{center}
\begin{tabular}{lcccccc}
\hline
\hline
Dwarf            &       & $1-B/A$ &       &       & $A_2$ &       \\
galaxy           & whole & inner   & outer & whole & inner & outer \\
\hline
Fornax           & 0.18  & 0.09    & 0.21  & 0.12  & 0.07  & {\bf 0.22}  \\
Sculptor         & 0.14  & 0.07    & 0.17  & 0.11  & 0.06  & 0.18  \\
Carina           & 0.20  & 0.17    & 0.21  & {\bf 0.20}  & 0.16  & {\bf 0.24}  \\
Ursa Minor       & 0.38  & 0.34    & 0.39  & {\bf 0.36}  & {\bf 0.29}  & {\bf 0.46}  \\
Sextans          & 0.07  & 0.09    & 0.07  & 0.08  & 0.08  & 0.08  \\
Draco            & 0.16  & 0.15    & 0.16  & 0.15  & 0.14  & 0.18  \\
Leo I            & 0.23  & 0.08    & 0.27  & 0.14  & 0.04  & {\bf 0.26}  \\
Leo II           & 0.09  & 0.10    & 0.09  & 0.09  & 0.10  & 0.09  \\
Sagittarius      & 0.47  & 0.32    & 0.51  & {\bf 0.39}  & {\bf 0.26}  & {\bf 0.61}  \\
LMC              & 0.07  & 0.19    & 0.04  & 0.15  & {\bf 0.20} & 0.08  \\
SMC              & 0.18  & 0.13    & 0.19  & 0.14  & 0.10  & {\bf 0.21}  \\
\hline
\end{tabular}
\tablecomments{For each dwarf galaxy
we list three values of parameters  $1-B/A$ and $A_2$: for the whole stellar sample within $0 < r < 2 r_{1/2}$,
for the inner sample within $0 < r < r_{1/2}$ and for the outer sample within $r_{1/2} < r < 2 r_{1/2}$. The values
of $A_2 \ge 0.2$ signifying the presence of bar-like shape were boldfaced.}
\end{center}
\end{table*}

Let us now focus on the 2D shapes of the simulated dwarfs, as they would be seen by a distant observer. We illustrate
the evolution of such shapes using our default simulation O1. In Figure~\ref{viewsrun1apo} we show how the dwarf's
stellar component would appear when viewed along the principal axes of the stellar distribution $x$, $y$ and $z$
(from the upper to the lower row) at subsequent apocenters of its orbit around the Milky Way (from the left to the
right column). The images were produced by selecting stars within $r<2$ kpc and rotating their positions to align
them with the principal axes determined for stars with $r < 2 r_{1/2}$, as described above. The plots show the surface
density distribution of the stars with contours equally spaced in log density. In agreement with the
conclusions from Figure~\ref{shapesrun1}, at second apocenter the dwarf already has a pronounced bar and its shape
changes from the oblate (initial disk) to the prolate (a bar). At the third apocenter the shape is again oblate
(as shown by the green line in the upper panel of Figure~\ref{cbbaapo} which crosses the diagonal line back again
from the prolate to the oblate part of the diagram at this time). Later on, the shape gradually becomes more and
more spherical with all axis ratios approaching unity. The final image of the dwarf, at the end of the simulation,
at its most non-spherical appearance (i.e., viewed along the $y$ axis) is shown in Figure~\ref{shapealongyrun1}.

In Figure~\ref{shapesalongy} we show analogous images for the remaining simulations O2-O7 and S6-S17. In all cases
stars within $r<2$ kpc were extracted from the last output of the simulation and rotated to align the distribution
with the principal axes. Only the views along the $y$ axis are shown, i.e. we see the dwarfs along the line of sight
in which they appear most non-spherical. The dwarfs are shown at the end of their evolution when most are far from
their pericenters, and therefore weakly affected by tidal forces at this time, except for the case of O4 which
has just passed its 6th pericenter. Although this dwarf has already reached an almost spherical shape during its
earlier evolution, it is strongly affected by tidal forces from the Milky Way at this particular instant (note that
its pericentric distance was 12 kpc at this passage) which manifests itself by strong distortion of the outer contours.

Going back to our single-number measures of the shapes, $1-B/A$ and $A_2$, we summarize our findings by comparing in
Figure~\ref{losshapes} the values of these parameters measured along the three lines of sight at the end of each
simulation and listed in columns 7-12 of Table~\ref{properties}.
For each run we plot $1-B/A$ with filled symbols and $A_2$ with open symbols. The triangles, squares
and circles correspond to measurements along the $y$, $x$ and $z$ axis of the stellar component, respectively.
As discussed above,
the observation along the intermediate axis $y$ always gives the most non-spherical appearance, i.e., triangles always
mark the highest values of $1-B/A$ and $A_2$. In addition, the next highest value of ellipticity is almost always the
one seen along the $x$ axis (so that the points are ordered triangle-square-circle from top to bottom). The exceptions
occur for runs O4, S7, S11 and S12. Since these cases correspond to $b/a < c/b$ this means that only in these four
cases do the dwarfs end up more prolate than oblate (O4 should be excluded because its shape is due to strong tidal forces
at pericenter). It is interesting, therefore, that although almost all dwarfs go through a bar-like phase during
their evolution, most of them end up more oblate than prolate.

\begin{figure}
\begin{center}
    \leavevmode
    \epsfxsize=7cm
    \epsfbox[10 10 215 420]{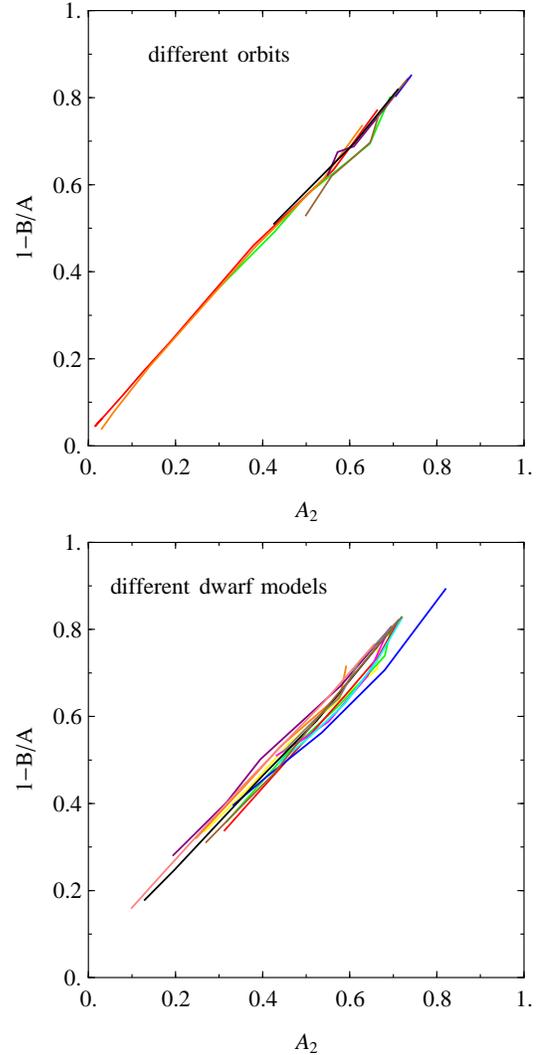}
\end{center}
\caption{The correlation between different measures of shape, $1-B/A$ and $A_2$, for the line of sight along the
$y$ axis (when the dwarfs appear most non-spherical) for dwarfs on different orbits O1-O7 (upper panel)
and dwarfs with different initial structure S6-S17 (lower panel). Color lines join measurements
at subsequent apocenters starting from the initial values at the upper right region of each panel.}
\label{a2baapo}
\end{figure}

We also note that the values of $1-B/A$ and $A_2$ are quite similar, although not exactly equal, and they preserve
the relative triangle-square-circle hierarchy of the measurements. To illustrate this similarity further, in
Figure~\ref{a2baapo} we plot $1-B/A$ as a function of $A_2$ for the line of sight where the dwarfs appear most non-spherical
(observation along $y$).
The lines join the values measured at subsequent apocenters, from the most non-spherical at the upper right corner
of each panel to the almost spherical in the lower left. The upper panel shows the results for simulations of dwarfs on
different orbits (O1-O7) and the lower one those with different structure of the dwarf (S6-S17). The lines corresponding
to each simulation can be identified by referring to the last column of Table~\ref{properties}. We can see that there
is a one-to-one correspondence between the two measures of shape, and thus conclude that both measures, although defined
differently from a mathematical point of view, essentially measure the departures from sphericity in a similar
way and can be used interchangeably to quantify the shapes of galaxies with resolved stellar populations.

\section{The shapes of Milky Way satellites}

In this section we discuss the shape parameters of the real dwarfs, satellites of the Milky Way using
photometric data. Our dwarf galaxy sample includes objects within heliocentric distance of about 260 kpc
(close to the largest apocenter considered in the simulations),
brighter than $M_V=-8$ mag and with good photometric data available. These requirements restrict
the sample to the eight classical dSph galaxies, the Sagittarius dwarf and the Large and Small Magellanic Clouds
(LMC and SMC). The data sets we used comprise the positions and magnitudes of stars in at least two bands to allow
for the selection of probable members of dwarf galaxies from the color-magnitude diagrams (CMD). In the case of the Fornax
dwarf we used the data from Coleman et al. (2005), for Sextans and Draco the data were retrieved from the
Sloan Digital Sky Survey database. For these dwarfs we selected member stars by iterative fitting of the position
of the red giant branch (RGB) and choosing stars within a strip of variable width
along the branch whose shape was adjusted to match
the equal-density contours of the stars in the CMD. For the LMC and SMC, we extracted the stars from the 2MASS survey
and also included stars from the asymptotic giant branch (see van de Marel
\& Cioni 2001; Gonidakis et al. 2009).

For the remaining dwarfs (Sculptor, Carina, Ursa Minor, Leo I, Leo II and Sagittarius) we used samples
obtained previously by S.R.M. and collaborators with more precise photometric membership determination using
color-magnitude as well as color-color diagrams to extract only the giant stars. The technique, described in detail
in Majewski et al. (2000a), relies on the sensitivity of the $DDO51$ filter to magnesium features that reflect stellar
surface gravity, temperature and abundance in later type stars. When combined with the
wide-band $M$ and $T_2$ filters of the Washington system, the $DDO51$ filter is especially useful for discriminating giant
stars from foreground dwarfs on the basis of differences in their respective $M-DDO51$ colors at a given $M-T_2$ color.
The data, although not publicly available, are described in the references provided in Table~\ref{dwarfs} and we
refer the reader there for details concerning each dwarf galaxy.

Table~\ref{dwarfs} lists the properties of the dwarfs included in this analysis: the dwarf name (column 1),
the position of the adopted center (columns 2 and 3), the heliocentric distance of the dwarf (column 4), the 2D
half-light radius $r_{1/2}$ in arcmin and kpc (columns 5 and 6), the number of member stars within $2 r_{1/2}$
(column 7), the method used for the membership determination (column 8, CM =  color-magnitude,
CM+CC = color-magnitude plus color-color), and the source of data (column 9, the three numbers give respectively
the reference for the dwarf center, distance and the photometric data used for the measurements of its shape).

Although the values of the half-light radii are available in the literature, for consistency we
determined them from the surface density distribution of the selected member stars by fitting a Plummer profile.
For this purpose, the surface density profiles were measured in logarithmic bins of projected radii and fitted with
data points weighted by density adjusting two parameters, the half-light radius and normalization.
The obtained values of $r_{1/2}$, listed in Table~\ref{dwarfs}, are in good agreement with published scales
(see, e.g., Table 2 in {\L}okas et al. 2011).

We measured the shape parameters $1 - B/A$ and $A_2$ first using member stars in the whole range $0 < r < 2 r_{1/2}$.
The values of these parameters are listed in the second and fifth column of Table~\ref{shapeparameters} respectively
and marked as `whole'. In order to distinguish between intrinsic bars and external tidal extensions
it is also interesting to look at the parameters calculated separately in the inner part within $0 < r < r_{1/2}$
and the outer part within $r_{1/2} < r < 2 r_{1/2}$. Those are shown in the columns of Table~\ref{shapeparameters}
marked respectively as `inner' and `outer'. In the following we will focus on $A_2$, which is a more customary
quantity used to describe bars, and adopt $A_2 \ge 0.2$ as a threshold signifying the presence of a bar-like shape.
Values above this threshold were typed in boldface in Table~\ref{shapeparameters}.

The motivation behind this division into inner and outer sample is that the tidal effects on the
shape of dwarf galaxies can take
two forms. The shapes can be affected by tides so that the whole orbital structure of the dwarf is rebuilt and a bar
forms, or only the outer contours of the stellar distribution are elongated signifying the transition to the tidal tails.
The latter situation is illustrated very well by the special case of simulation O4 which ends at the pericenter of the orbit.
This moment of the evolution is shown in the upper right panel of Figure~\ref{shapesalongy}. While the inner contours of
the stellar density distribution are circular, the outer ones are strongly elongated due to strong tidal force at the pericenter
of the orbit. On this orbit with a very small pericenter (as well as for the very tight orbit O2)
the dwarf transforms into an almost perfectly spherical shape early on in the evolution, as demonstrated by the
time-dependence of the shape parameters in Figures~\ref{axisratios2-19}-\ref{a2xyz2-19}. The shape remains spherical for
most of the time at later stages, except at pericenters, where strong elongation takes place.

\begin{figure*}
\begin{center}
    \leavevmode
    \epsfxsize=16cm
    \epsfbox[140 0 560 560]{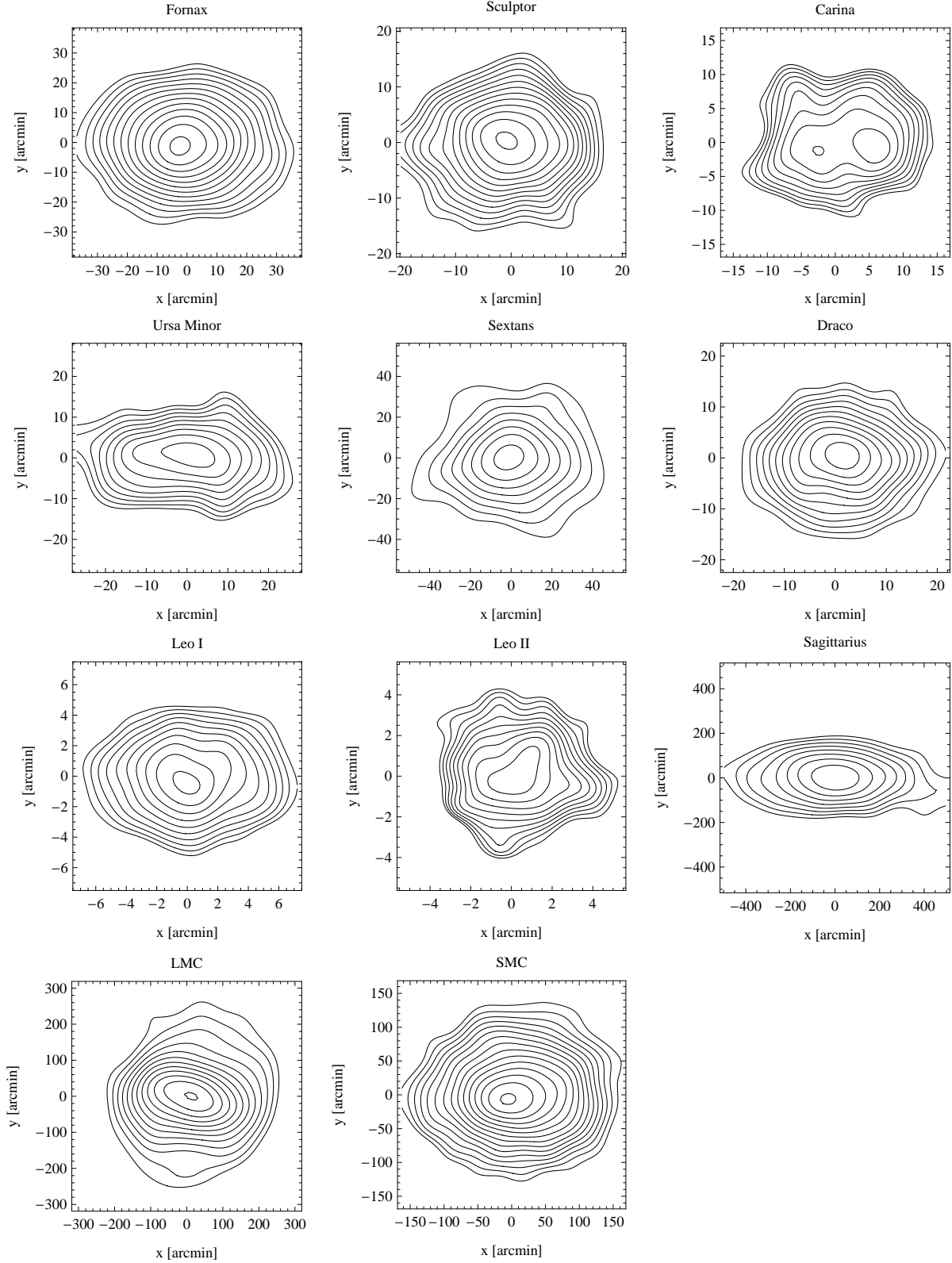}
\end{center}
\caption{The surface density distribution of stars in Milky Way satellites. In all cases
(except for the LMC and SMC, where the number
of stars is large) the surface density of stars was smoothed with Gaussian filters of scale $0.2 r_{1/2}$ (where $r_{1/2}$
is the half-light radius from Table~\ref{dwarfs}) and the maps were created by binning the resulting stellar
distribution into a $10 \times 10$ mesh. The contours are equally spaced in log surface density distribution.
The images were rotated to align
the major axis (as determined within $2 r_{1/2}$) with the $x$ axis of the plots.}
\label{shapessmoothed}
\end{figure*}

In Figure~\ref{shapessmoothed} we show the contour maps of the surface density of the stars in real dwarfs up to
projected radii of the order of $2 r_{1/2}$. The maps were obtained by smoothing the discrete distribution of stars
with Gaussians of scale $\sigma = 0.2 r_{1/2}$ and binning it into $10 \times 10$ grids. As discussed in section~3,
such smoothing scales are small enough not to obscure subtle morphological features.
The contours joining the regions of equal surface density of the stars are equally spaced in log surface density.

As verified in Table~\ref{shapeparameters}, only in the case of Ursa Minor and Sagittarius are all three values of the bar
mode parameter --- for the whole, inner and outer samples --- found to have $A_2 > 0.2$.
This strong elongation, determined by the
discrete shape measures, is confirmed in a more qualitative way by the images of these dwarfs in Figure~\ref{shapessmoothed}.
We conclude that these dwarfs possess the
tidally induced bar and that their outer parts are affected by tides. In the case of the Sagittarius dwarf, the outer parts
are likely stretched by tidal forces acting {\em now\/}, since the dwarf is very close to the pericenter of the orbit
and the pericenter distance is very small ($<15$ kpc); (like the dwarf in simulation
O4 at the end of the evolution). In the case of Ursa Minor, which is much farther away from the center of the Milky Way at
present, the outer contours probably signify a transition to the tidal tails.
For the remaining dwarfs, the inner bar is detected only in the LMC and possibly in Carina ($A_2 = 0.16$ in the inner part).
In addition, Fornax, Carina, Leo I
and the SMC have $A_2 > 0.2$ in the outer parts, signifying mild tidal extensions.

\section{Discussion}

We have studied the shapes of dwarf galaxies orbiting the Milky Way using new discrete measures appropriate for systems
where individual stars are resolved. First, we traced the evolution of shape parameters in a set of collisionless
$N$-body simulations of late type progenitors of present day dSph galaxies. The initially disky dwarfs were placed
on different orbits around a Milky Way-type galaxy and the inner structure of the satellites was also varied. In addition to the
strong mass loss and randomization of stellar orbits, tidal interactions inherent to such configurations lead to
significant evolution of the shape. Usually at the first pericenter passage the dwarfs undergo bar instability and
their overall distribution of stars transforms from the initial disks to triaxial bars. The bars are preserved for a few
Gyrs of evolution but become shorter in time and produce, in the most tidally affected cases, almost spherical
stellar components.

The effect of tidal forces on the shape of the dwarfs can be twofold. In addition to the tidally induced bar, which can
be easily identified with the full 3D information available in the simulations, the tidal forces can stretch the outer
contours of the stellar distribution even in dwarfs that have already evolved into spheres. This is particularly well
visible at pericenters but can also occur at other parts of the orbit where the dwarf is not being strongly shocked, but
is already surrounded by the material stripped from its main body in the form of tidal tails. The two effects can
be distinguished by measuring shapes separately in the inner and outer parts of the dwarfs.

We found the bar phase of the evolution to be quite common among the simulated dwarfs. Given that the dwarf models were
not susceptible to bar instability in isolation, the presence of the bars can be considered as a signature of tidal stirring.
One has to keep in mind however, that the tidal forces are extremely effective not only in inducing the bars but also in
destroying them. If the tidal force is very strong, the bar can transform into a spherical shape over a rather short timescale.
It would therefore be very difficult to find a strong correlation between the formation of a bar and the strength of
the tidal force or a dwarf's distance from the Milky Way.

Nevertheless, given the variety of shapes within the dwarf galaxy population in the vicinity of the Milky Way we attempted
to check if the tidal
stirring scenario is broadly consistent with the properties of real dwarfs. To this end we measured the shapes of
brighter dwarf galaxies having good enough available photometry for large samples of stars. Out of eleven dwarfs studied,
we found clear signatures of inner bars in Ursa Minor, Sagittarius, LMC and possibly Carina. In six out of eleven dwarfs
we found indications of tidal extensions in the outer parts which likely signify transitions to the tidal tails.

The elongated
shapes of Ursa Minor, Sagittarius, Carina and the inner bar of the
LMC have been pointed out by many previous studies. The results
are however particularly interesting for Sagittarius because its
elongated shape is usually interpreted as due to tidal forces
acting at the present time when the galaxy is at the pericenter of its orbit.
We find instead that the shape is elongated down to the
very center of the dwarf thus lending support to the model proposed by {\L}okas et al. (2010a). In this scenario, the
progenitor of Sagittarius was a disk galaxy similar to the present LMC, which transformed into a bar at the first pericenter
passage. The bar is preserved until the present and the dwarf is currently at the second pericenter on its orbit around the
Milky Way. The presence of the inner bar in the LMC also seems consistent with the tidal stirring scenario which predicts that
bars are typically formed at the first pericenter. Since the LMC is probably on its first passage around the Milky Way
(Besla et al. 2007), it should be forming its bar right now.

It may seem surprising that among the eight classical dSph galaxies in the vicinity of the Milky Way a clear inner bar is
detected only in Ursa Minor which, according to the scenario presented here, would imply that it has not passed many
pericenters on its orbit around the Milky Way. On the other hand, Ursa Minor is dominated by the most metal-poor
stellar population among the classical Milky Way dSphs (Kirby et al. 2011). This means that it lost its gas rather
early, probably via ram pressure stripping on its first passage around the Milky Way (Mayer et al. 2007), a mechanism
acting most effectively for dwarfs accreted early. However, the gas may also have been lost via a strong blow-out in
the formation of the initial stellar population while the system was still evolving in isolation. Given the
stochasticity in the star formation history outcomes, as well as the wide distribution of orbital
and structural parameters of the progenitor dwarfs, it would be difficult to exactly reproduce the history of any
given dwarf just from its present shape. The case of Ursa Minor however proves that bars are present even among the eight
classical dwarfs and may be present also in other systems but not detectable because of their orientation along
our line of sight.

The present study was based on photometric data alone. However, considerable progress towards discriminating
between different scenarios of dSphs formation is expected if these data are combined with kinematic measurements.
Although rich samples of radial velocity measurements are presently available only for a few dSphs (e.g., Walker
et al. 2009), such samples are bound to grow substantially in the near future thanks to a new generation of multi-object
spectrographs (such as the Prime Focus Spectrograph on the Subaru telescope). Kinematical samples for a few thousand
stars could help detect remnant rotation that should be present in at least some dSphs if they
formed from disky progenitors, as envisioned by the tidal stirring model. Such a rotation signal
has until now only been tentatively detected for some dwarfs. In addition, such data will also help distinguish
bars from other structures. Bars seen perpendicular to the longest axis should show little rotation in the inner parts,
contrary to disks seen edge on. On the other hand, rotation detected in systems that appear spherical may indicate
a bar seen along the longest axis (see {\L}okas et al. 2010b).

We end with a note of caution that the picture established here with the help of collisionless simulations using standard
assumptions concerning, e.g., the initial dark matter profiles of the dwarfs, may well be modified if hydrodynamical
processes are included or the assumptions changed. Although few hydrodynamical simulations of tidal
stirring have been performed
to date, they indicate that the shapes may be significantly affected by the presence of the gas and star formation
(Mayer et al. 2007). The details of the morphological transformation will also be different if the initial dark matter
profile has a core rather than a cusp (Governato et al. 2010; {\L}okas et al. 2012) and the initial stellar and
gaseous disk of the dwarf is thicker. We plan to address these issues in our forthcoming papers.

\section*{Acknowledgments}

This research was partially supported by the Polish National Science Centre under grant N N203 580940.
We wish to thank Gary Da Costa for providing photometric data for the Fornax dwarf in electronic form.
S.R.M. is grateful for financial support from NSF grants AST 97-02521, AST 03-07851, AST 03-07417
and AST 08-07945, a fellowship from the David and Lucile Packard Foundation, and a Cottrell Scholarship
from The Research Corporation. He also thanks William E. Kunkel, James C. Ostheimer, Christopher Palma,
Richard J. Patterson, Michael H. Siegel, Sangmo Tony Sohn, and Kyle B. Westfall for their help in producing
the photometric catalogs used in this analysis.
S.K. is funded by the Center for Cosmology and Astro-Particle Physics (CCAPP) at The Ohio State University.
J.L.C. acknowledges support from National Science Foundation grant AST 09-37523.
The work of L.A.M. was carried out at the Jet Propulsion Laboratory, under a contract
with NASA. L.A.M. acknowledges support from the NASA ATFP program.
The numerical simulations were performed on the Cosmos cluster at the Jet Propulsion Laboratory.
This work also benefited from an allocation of computing time from the Ohio Supercomputer
Center (http://www.osc.edu).

\end{document}